\documentclass[superscriptaddress,aps,journal=prl,preprint]{revtex4-2}
\usepackage{graphicx}
\usepackage{makecell, multirow}
\usepackage{color, bm, soul, xcolor}
\usepackage{amsmath, amssymb}
\usepackage{geometry}
\usepackage{braket}
\usepackage[colorlinks=true, linkcolor=black, citecolor=black, urlcolor=blue]{hyperref}
\newcommand{\eg}{{\em e.\,g.}}

\newcommand{\etal}{{\em et al}.\ }

\begin{document}

\title{Topological Control of Transition Metal Networks for Reversible High-Capacity Li-rich Cathodes}


\author{Changming Ke}
\altaffiliation{These two authors contributed equally}
\affiliation{Department of Physics, School of Science, Westlake University, Hangzhou, Zhejiang 310030, China}
\affiliation{Institute of Natural Sciences, Westlake Institute for Advanced Study, Hangzhou, Zhejiang 310024, China}

\author{Yudi Yang}
\altaffiliation{These two authors contributed equally}
\affiliation{Department of Physics, School of Science, Westlake University, Hangzhou, Zhejiang 310030, China}
\affiliation{Institute of Natural Sciences, Westlake Institute for Advanced Study, Hangzhou, Zhejiang 310024, China}

\author{Minjun Wang}
\affiliation{Institute of Zhejiang University-Quzhou, Zheda Road 99, Quzhou 324000, China}

\author{Jianhui Wang}
\email{wangjianhui@westlake.edu.cn}
\affiliation{Zhejiang Key Laboratory of 3D Micro/Nano Fabrication and Characterization, Department of Electronic and Information Engineering, School of Engineering, Westlake University, Hangzhou 310030, China.}
\affiliation{Division of Solar Energy Conversion and Catalysis at Westlake University, Zhejiang Baima Lake Laboratory Co., Ltd. Hangzhou 310000, China}
\affiliation{Research Center for Industries of the Future (RCIF), Westlake University, Hangzhou 310030, China}

\author{Shi Liu}
\email{liushi@westlake.edu.cn}
\affiliation{Department of Physics, School of Science, Westlake University, Hangzhou, Zhejiang 310030, China}
\affiliation{Institute of Natural Sciences, Westlake Institute for Advanced Study, Hangzhou, Zhejiang 310024, China}

\begin{abstract}
Developing high-energy-density batteries is essential for advancing sustainable energy technologies. However, leading cathode materials such as Li-rich oxides, including Li$_2$MnO$_3$, suffer from capacity loss due to irreversible oxygen release and structural degradation, both consequences of the oxygen redox activity that also enables their high capacity. The atomic-scale mechanisms behind this degradation, and whether it can be made reversible, remain open questions. Here, using submicrosecond-scale molecular dynamics simulations with first-principles accuracy, we directly visualize the entire charge–discharge cycle of Li$_2$MnO$_3$, uncovering the full lifecycle of the O$_2$-filled nanovoids responsible for degradation and identifying the critical size limit for voids to remain fully repairable upon discharge. Our results reveal that the topology of the Mn cation network is the key factor governing void growth, coalescence, and reparability. Based on a structural topology-informed design principle, we computationally develop a novel Li$_2$MnO$_3$ structure featuring a Mn lattice with a Kagome-like pattern, demonstrating full electrochemical reversibility even under extreme 80\% delithiation. 
Our work establishes a new paradigm for designing high-energy cathodes, shifting the focus from mitigating damage to engineering inherent stability through atomic-level topological control of transition metal network.
\end{abstract}
\maketitle
\newpage

\section{Introduction}

The pursuit of higher energy density in lithium-ion batteries has brought Li-rich Mn-based oxides, such as $x$Li$_{2}$MnO$_{3}{\cdot}(1-x)$Li$M$O$_{2}$ ($M$=Ni, Co, Mn), into focus as highly promising next-generation cathode materials~\cite{Li18p1800561, Poizot20p6490, Assat18p373, Li25p2500940, Zhang22p522, Wang24p2405659}. Their distinct advantage lies in the Li$_{2}$MnO$_{3}$ component, which can unlock capacities exceeding those of conventional cathodes by activating anionic redox, a process where lattice oxygen participates in charge compensation~\cite{Rana20p634, Li17p1701054, Zhang22p522,  Assat18p373, Sathiya13p827, Li18p1704864, Seo16p692, Li18p1705197, Luo16p684, Ben19p496, Qiu25p941}. However, this high-capacity mechanism comes at a cost. Oxygen redox, typically activated above 4.6 V, leads to lattice oxygen loss~\cite{Zhang22p5641, Csernica25p92, House20p777, Wang24p12343}, triggering a cascade of detrimental effects, including irreversible phase transformations~\cite{Zhang22p5641, Mohanty13p6249, Hong10p10179, Wang24p2313672}, particle cracking, and severe capacity and voltage degradation~\cite{Li25p3441, He21p2005937, Li18p1800561, Zhao21p716}. This fundamental instability remains a major obstacle to the practical deployment of these high-capacity materials.

While the link between oxygen release and performance degradation is well established~\cite{Zhang22p5641, Li25p3441, Gao25p743}, the atomic-level mechanisms governing this process have been the subject of intense debate. A critical question is whether the molecular O$_{2}$ formed during charging is an inherently irreversible byproduct, or if it can be exploited for reversible capacity. Seminal studies by House and co-workers suggested that molecular O$_{2}$ trapped within bulk nanovoids could be reversibly reduced, contributing to capacity~\cite{House20p777, House21p2975, House21p781}. In contrast, other reports have associated capacity loss with O$_2$ that either escapes from surfaces or becomes permanently trapped in the structure~\cite{Marie24p818}. Despite these insights, a fundamental understanding of the factors that determine the fate of O$_{2}$, whether reversible or irreversible, remains elusive. Moreover, the atomistic mechanisms governing the formation, growth, and potential healing of the O$_{2}$-hosting voids are still poorly understood.

Resolving this requires direct, atomic-scale visualization of the entire charge--discharge process. Phenomena such as mechanical strain and dislocation formation, often cited as concurrent degradation mechanisms~\cite{Singer18p641, Yan18p2437, Liu22p305}, are likely manifestations of underlying atomic reorganization driven by oxygen activity. However, a critical ``observational gap'' has hindered progress. Experimental techniques lack the combined spatial and temporal resolution to track individual atomic motions over the nanoseconds to microseconds required for void evolution~\cite{Song22p1049}. Conversely, while density functional theory (DFT) provides high-fidelity atomic insights, its prohibitive computational cost limits simulations to picosecond timescales~\cite{Anisimov97p767, Timrov22p033003, Timrov23p9061, Mccoll24p826, Seo16p692}, far too short to capture the crucial dynamics of void growth and repair. This challenge is compounded by dynamic charge transfer between Mn and O ions during electrochemical cycling, which introduces complex, coupled electronic and structural transformations. As a result, accurately modeling the degradation pathways, particularly those involving oxygen redox, remains a significant challenge, hindering predictive design of stable, high-capacity cathode materials.

This work bridges the critical observational gap by employing submicrosecond-scale molecular dynamics (MD) simulations, enabled by a machine learning deep potential, to achieve first-principles accuracy over unprecedented timescales. 
We present a comprehensive model of the full charge–discharge cycle of Li$_{2}$MnO$_{3}$, capturing the entire lifecycle of O$_{2}$-filled nanovoids. Our simulations reveal a self-catalytic process: molecular O$_{2}$ promotes the conversion of additional lattice oxygen, thereby promoting void growth. This process is accompanied by a striking dichotomy in Mn diffusion. Stabilizing \textit{intralayer} migration forms condensed triangular Mn networks that act as ``walls'' to confine void growth. In contrast, stochastic \textit{interlayer} migration into adjacent Li layers disrupts this containment, leading to uncontrolled void expansion. This divergence in migration behavior of Mn ultimately determines the size of the nanovoids, which is key to reversibility. We find nanovoids smaller than 1~nm allow trapped O$_{2}$ to be fully reincorporated into the lattice upon discharge, while larger voids become irreparable. Based on these insights, we propose a design principle: pre-engineering an Mn topology composed of a weaved triangular lattice that defines a uniform array of repairable, sub-nanometer voids. As a proof of concept, we designed a novel Li$_{2}$MnO$_{3}$ structure with a Kagome-like Mn topology. In our simulations, this design confines O$_{2}$ within 1-nm voids even at 80\% delithiation, yielding a 20\% increase in fully reversible theoretical capacity. The topology-centric strategy shifts the focus from mitigating degradation to designing inherent reversibility, offering a new path toward stable, high-energy cathodes.

\section{Results and Discussion}
\subsection{Simulating charging process via Li extraction}
\begin{figure}[htb]
\centering
\includegraphics[width=17 cm]{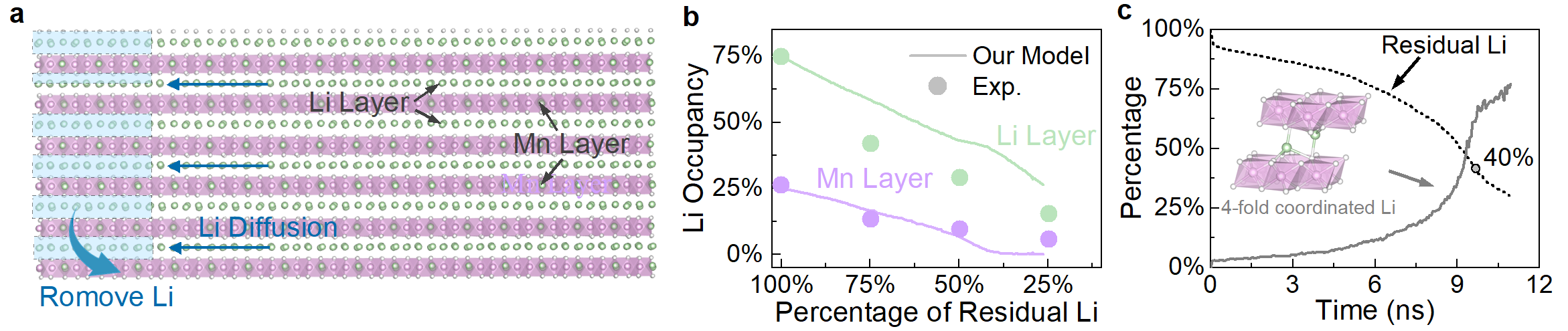}
\caption{Molecular dynamics simulation of the charging process. (a) Schematic illustration of the simulation setup, where Li atoms (Li$^+$ + e$^-$) are sequentially removed from a designated region of the Li layers within the supercell. 
Electroneutrality is maintained by assuming that Li$^+$ removal occurs instantaneously relative to the slower timescale of structural relaxation. (b) 
Simulated Li occupancies in the Li and Mn layers (lines) 
at various states of delithiation  show excellent agreement with experimental results (solid circles) from Li \etal~\cite{Liu16p1502143}. (c) 
Time evolution of 4-fold coordinated Li over a 10 ns MD trajectory. 
Delithiation induces a transition from 6-fold to 4-fold coordination, forming a dumbbell structural motif (inset).}
  \label{fig_removeMD}
\end{figure}

To investigate the atomistic dynamics of Li$_x$MnO$_3$ ($0.5\leq x \leq 2.0$) during electrochemical cycling, we developed a deep potential (DP) model trained on DFT-calculated energies and atomic forces for a comprehensive set of configurations (see details in the Supplementary Material). This approach enables first-principles-accurate, atomic-resolution modeling over submicrosecond timescales necessary to capture key structural evolution processes. To mimic the localized delithiation fronts that occur in real electrodes during charging, neutral Li atoms (Li$^+$ + e$^-$) were progressively removed from specific regions within the Li layer of the supercell (Fig.~\ref{fig_removeMD}a). 
The accuracy and predictive capability of this computational framework are validated by its strong agreement with key experimental observations.

First, our model accurately captures the simultaneous extraction of Li from both the pure Li layer and the Mn-containing layer, a characteristic feature of Li-rich cathodes. As shown in Fig.~\ref{fig_removeMD}b, the simulated Li occupancies in the Li and Mn layers, plotted as a function of the residual Li fraction, show excellent agreement with experimental measurements~\cite{Liu16p1502143}. Second, our simulations capture the complex evolution of Li-ion coordination with oxygen within the lattice. 
As charging proceeds, the remaining Li ions migrate from their initial 6-fold octahedral coordination to more stable 4-fold tetrahedral sites. This transition is reflected by the increasing population of 4-fold coordinated Li ions, as shown in Fig.~\ref{fig_removeMD}c. Notably, our simulation reveals the spontaneous self-assembly of an ordered sublattice characterized by Li-Li dumbbells, where Li ions occupy adjacent tetrahedral sites (Fig.~\ref{fig_removeMD}c inset). This unique structural motif was recently identified experimentally by Song \etal\cite{Song25p23814}, providing atomic-scale validation of our model's accuracy. The ability to reproduce both the macroscopic compositional changes in Li and Mn layers, as well as the specific atomic rearrangements, lends strong confidence to the fidelity of our model, establishing it as a reliable tool for investigating the more elusive mechanisms of oxygen redox and degradation.

\subsection{Mn Disordering and void formation}

\begin{figure}[htb]
\centering
\includegraphics[width=17.5 cm]{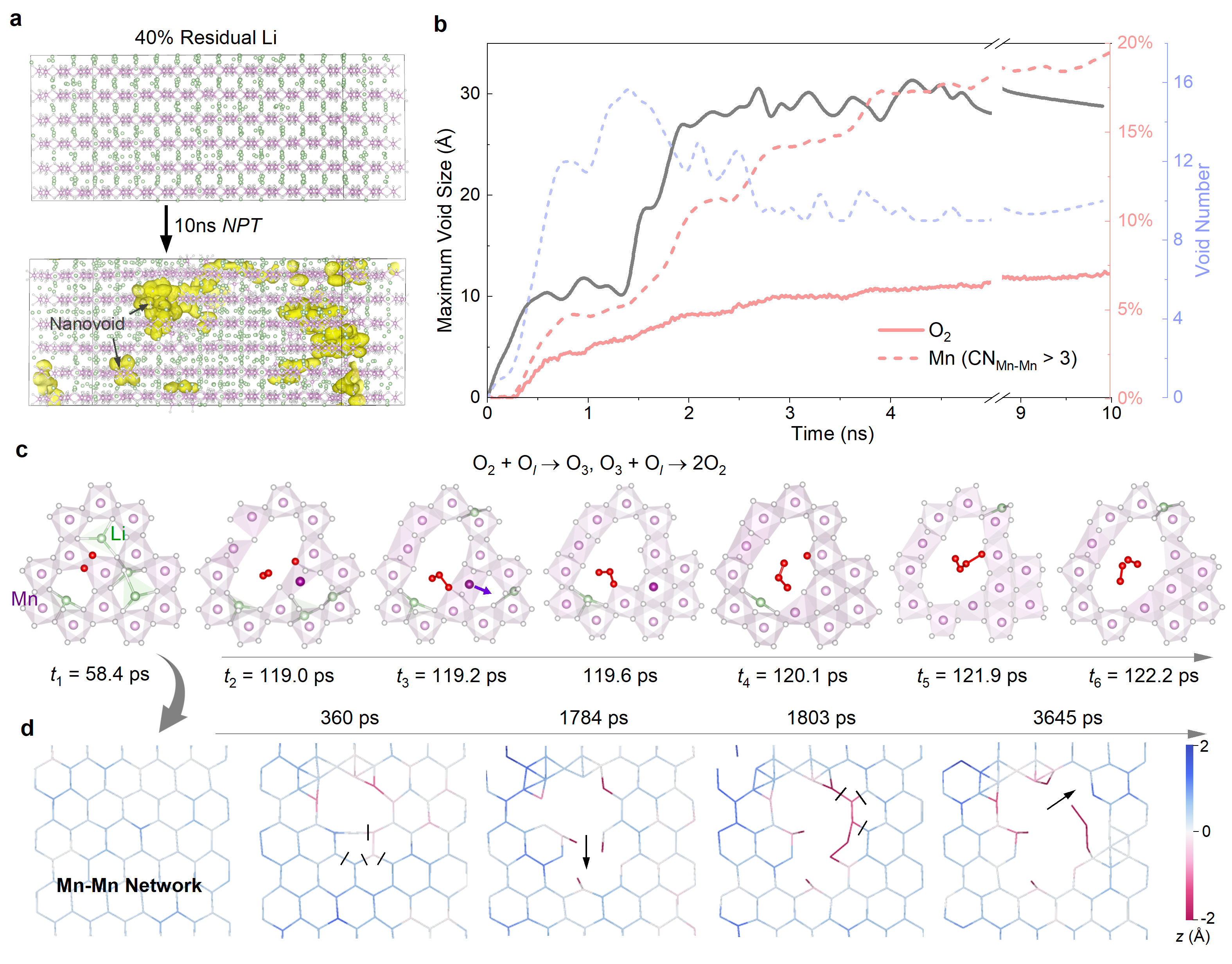}
 \caption{
 Atomistic mechanism of nanovoid nucleation and growth in the 40\% residual Li structure.
(a) Snapshots from MD simulations showing nanovoids (yellow isosurfaces) with trapped O$_2$ molecules distributed heterogeneously throughout the structure, taken after 10 ns of simulation at 500 K.
(b) Time evolution of key structural metrics. The size and number of nanovoids initially increase simultaneously, accompanied by increased Mn disorder (CN$_{\text{Mn–Mn}} > 3$) and rising O$_2$ content. The rapid increase in void size and the simultaneous decrease in void number around 1.4 ns are caused by void coalescence.
(c) Schematic illustration of the self-catalytic cycle for O$_2$ formation and the nanovoid nucleation mechanism derived from MD simulations.
(d) Mn atomic network showing nanovoid growth driven by interlayer Mn migration. The initial Mn framework adopts a honeycomb lattice, with atoms colored by their out-of-plane displacement ($z$) relative to the layer. Mn atoms with negative $z$ values (colored in red) indicate ongoing interlayer diffusion, which drives void growth.
}
\label{fig_04Li}
\end{figure}

Experimentally, charging Li-rich cathodes above 4.6~V is known to trigger O$_2$ release and the formation of structural voids~\cite{Zhang22p5641}. To uncover the atomistic origin of this degradation, we performed MD simulations on a Li$_{x}$MnO$_3$ supercell at 40\% residual Li, corresponding to a capacity of 275 mAh/g. After a 10~ns simulation at 500~K, the initially ordered structure evolves dramatically, forming extensive nanovoids filled with molecular O$_2$, as visualized in Fig.~\ref{fig_04Li}a. 
Figure~\ref{fig_04Li}b provides a time-resolved statistical analysis of this transformation. We track molecular O$_2$ formation by calculating the percentage of oxygen atoms with an O-O coordination number (CN$_{\rm O\text{-}O}$) greater than one (using a 1.6~\AA{} cutoff).
Mn disordering is quantified by the percentage of Mn atoms with a Mn-Mn coordination number greater than three (CN$_{\rm Mn\text{-}Mn}>3$), indicating a deviation from the pristine honeycomb lattice due to Mn clustering. 
The simultaneous increase in both molecular O$_2$ and Mn clustering, shown in Fig.~\ref{fig_04Li}a, suggests that these two processes are coupled.
Concurrently, the number of nanovoids rises, but after $\approx$1.4~ns, a sharp drop in  void number is accompanied by a rapid increase in maximum void size, signaling the onset of void coalescence. By 10~ns, the system reaches a quasi-equilibrium state where $\approx7$\% of lattice oxygen has converted to molecular O$_2$ and 20\% of Mn ions are disordered. This results in large, coalesced nanovoids reaching diameters up to 3~nm. As discussed later, nanovoids smaller than 1 nm containing trapped O$_2$ molecules can fully recover during discharge, whereas larger voids become irreparable.

This raises a critical question: what drives the system toward the formation of large, damaging  nanovoids rather than smaller, benign ones? Our atomic-scale analysis first reveals an self-catalytic mechanism for O$_2$ production that promotes localized void growth. As shown in Fig.~\ref{fig_04Li}c, during delithiation, under-coordinated lattice oxygen atoms (O$_l$) can stochastically come into close proximity and combine to form an initial O$_2$ molecule ($t_1$ and $t_2$ snapshots).
This O$_2$ can then react further with a nearby O$_l$ to form a transient O$_3$ intermediate with a life time of $\approx$1~ps (from $t_3$ to $t_4$), which subsequently reacts with another O$_l$ and rapidly decomposes into two new O$_2$ molecules: O$_2$ + O$_l$ $\rightarrow$ O$_3$, followed by O$_3$ + O$_l$ $\rightarrow$ 2O$_2$ (see $t_5$ and $t_6$). 
This chemical feedback loop is tightly coupled with a structural feedback loop: O$_2$ formation leaves behind under-coordinated Mn ions, which become mobile and migrate away from the reaction site (see $t_3$ and $t_4$). These migrating Mn ions reorganize into condensed, triangular motifs that form structural ``walls” around the growing voids (post-$t_5$ snapshots), consistent with the observed rise in Mn atoms with CN$_{\rm Mn-Mn} > 3$ in Fig.~\ref{fig_04Li}b. 
Trapped O$_2$ molecules further react with under-coordinated lattice oxygen atoms along the void boundaries. Collectively, these coupled feedback loops drive the localized buildup of O$_2$ and the heterogeneous void nucleation.

While the self-catalytic cycle explains the stochastic void nucleation, the formation of large cavities is governed by the evolving topology of the Mn atomic network.
We find that Mn migration exhibits a clear dichotomy. 
As shown in Fig.~\ref{fig_04Li}d, intralayer Mn migration leads to a local transition of Mn honeycomb lattice into triangular lattice that confines voids and suppress their in-plane expansion. This is because lattice oxygen atoms within triangular Mn frameworks are more fully coordinated and therefore less chemically reactive.
However, the structural stability of void boundary is compromised by stochastic interlayer (out-of-plane) Mn migration. As captured in the $z$-height color map of Fig.~\ref{fig_04Li}d, Mn atoms migrating into adjacent Li layers (also acquiring CN$_{\rm Mn\text{-}Mn} > 3$ ) show a color shift from blue (positive $z$) to red (negative $z$). 
Unlike their intralayer counterparts, these interlayer Mn ions fail to form compact triangular frameworks, leaving surrounding oxygen atoms under-coordinated and chemically active.
This exposure sustains the self-catalytic O$_2$ formation process described earlier, ultimately driving void growth. Notably, we observe that the coalescence of neighboring nanovoids is also closely associated with interlayer Mn migration, which breaches the relatively weak Mn honeycomb lattice at void boundaries.

\subsection{Void percolation and O$_2$ loss channel}

\begin{figure}[htb]
\centering
\includegraphics[width=8.5 cm, clip]{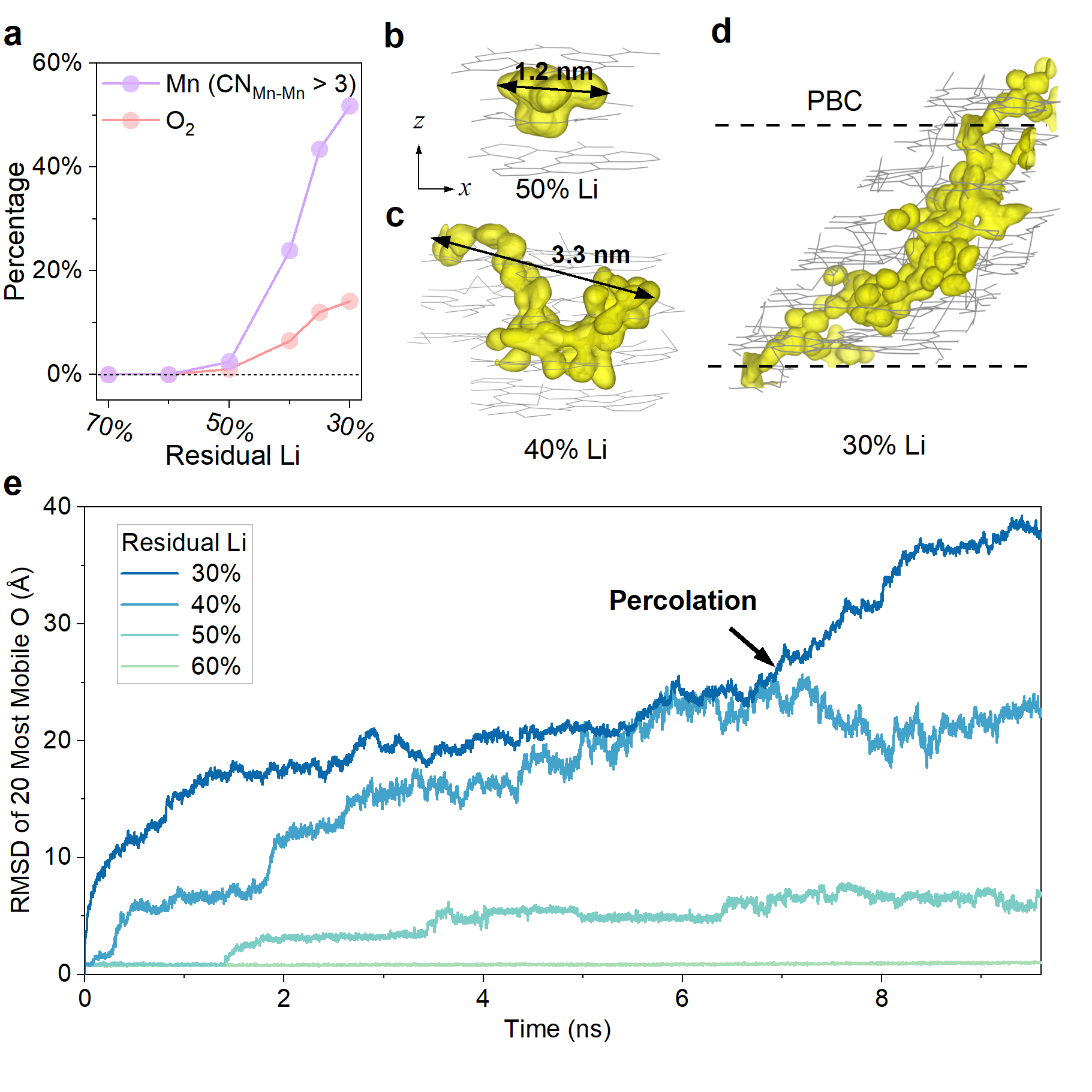}
 \caption{Nanovoid size and connectivity at various delithiation levels. (a) O$_2$ content and Mn disorder versus residual Li content. Both increase sharply as Li content decreases, indicating structural destabilization.
 (b–d) Nanovoids evolution with delithiation. Nanovoids increase in size as Li content decreases. At 30\% Li, they coalesce into a continuous diffusion channel spanning the simulation cell with periodic boundary condition (PBC).
 (e) Root-mean-square displacements (RMSD) of the 20 most mobile oxygen atoms at different residual Li contents.}
\label{fig_Licontent}
\end{figure}

Experimentally, deep delithiation of Li-rich cathodes  is known to trigger irreversible oxygen release~\cite{Zhang22p5641}, yet the physical pathway by which O$_2$ escapes from the bulk crystal into the electrolyte has remained elusive. By simulating the electrode at various charge states, We mapped a progression from isolated nanovoids to a percolating network of interconnected channels that ultimately enables long-range O$_2$ diffusion. As shown in Fig~\ref{fig_Licontent}a, both O$_2$ content and Mn disorder (featuring CN$_{\text{Mn-Mn}}>3$) remain negligible at moderate delithiation levels (\eg, $>$50\% residual Li). This corresponds to the formation of small, isolated nanovoids, such as the $\approx$1.2 nm nanovoid observed at 50\% Li (Fig.\ref{fig_Licontent}b). The limited mobility of O$_2$ in this regime is further evidenced by the low root-mean-square displacement (RMSD) of oxygen atoms at 60\% and 50\% residual Li content, as shown in Fig.~\ref{fig_Licontent}e.

However, as the residual Li content drops below a critical threshold, both O$_2$ formation and Mn disorder increase sharply  (Fig.~\ref{fig_Licontent}a). 
As charging proceeds to deeper levels (40\% Li), nanovoids expand significantly to $\approx$3 nm in diameter (Fig.~\ref{fig_Licontent}c), accompanied by a noticeable rise in oxygen mobility (Fig.~\ref{fig_Licontent}e). The critical structural transition occurs at $\approx$30\% residual Li: voids begin to coalesce into a continuous, percolated network that spans the simulation cell (Fig.~\ref{fig_Licontent}d). This percolation is marked by a sustained increase in oxygen RMSD over time (Fig.~\ref{fig_Licontent}e), indicating that O$_2$ molecules are no longer confined but can now diffuse freely over long distances. The formation of this interconnected channel network provides a direct diffusion pathway for molecular O$_2$ to exit the particle interior and reach the cathode-electrolyte interface, resulting in irreversible gas release. This mechanism offers a structural explanation for severe degradation phenomena observed experimentally, including damage to the solid electrolyte interphase, anode deterioration, and even thermal runaway under aggressive cycling conditions~\cite{Li25p3441}.

\subsection{Partial Repair of Voids During Discharging}

We next address a key question: to what extent can nanovoids be repaired during the discharge cycle if O$_2$ molecules remain trapped within the cathode?
This question is critical for understanding the limits of reversible cycling and for developing strategies to mitigate long-term degradation.
To explore this, we used DPMD to perform, likely for the first time, a full charge–discharge cycle computationally at atomic resolution (see additional details in Supplementary Material). As illustrated in Fig.~\ref{fig_discharge}a, we first simulated charging by removing Li to generate structures with 50\%, 40\%, and 35\% residual Li. Each structure was equilibrated for 10 ns to allow the formation of O$_2$-containing nanovoids. These equilibrated configurations then served as the starting point for discharge simulations, in which Li atoms (Li$^+$ + e$^-$) were progressively reinserted into the lattice.

The results, shown in the right panel of Fig.~\ref{fig_discharge}a, demonstrate that capacity reversibility is highly sensitive to the depth of the initial charge. When charged to 50\% residual Li, the structure recovers 99.9\% of its Li content upon discharge, indicating near-complete electrochemical and structural reversibility (Fig.~\ref{fig_discharge}b, top).
However, deeper charging significantly reduces recovery: capacity drops to 96\% for the 40\% residual Li structure and to just 90\% for the 35\% residual Li structure. These losses are accompanied by incomplete structural restoration and the persistence of irreversible voids (Fig.~\ref{fig_discharge}b, bottom), indicating permanent degradation and capacity loss.

The atomistic mechanisms underlying the observed differences in reversibility are illustrated in Figs.~\ref{fig_discharge}c-d. Under shallow charging conditions like 50\% residual Li, the resulting voids are small, approximately 1 nm in diameter (Fig.~\ref{fig_discharge}c, left). During discharge, these voids are fully healed. Reinserted Li ions react with the trapped O$_2$ molecules, initiating a redox process that reconstructs the local crystal lattice. This involves the transient formation of LiO$_2$-like species along the void walls. Critically, the small void size keeps Mn ions at the boundary in close proximity to the reaction front. These Mn ions actively participate in the redox process, providing electronic compensation and stabilizing the otherwise unstable LiO$_2$ intermediates. 

In contrast, the repair process fails for the larger $\approx$2.5 nm voids formed during deep charging (35\% residual Li), as shown in Fig.~\ref{fig_discharge}d. Although initial Li reinsertion leads to the formation of LiO$_2$ patches at the void perimeter, causing partial contraction, healing stalls once the void shrinks to about 1.5 nm. 
Without nearby Mn ions to sustain redox activity, O$_2$ molecules trapped in the void center remain chemically inert. Additionally, further healing would require the formation of extended LiO$_2$ domains, which are thermodynamically unstable in the absence of adjacent Mn coordination. As a result, the repair process halts, leaving behind a persistent, irrecoverable void that locks in structural damage. This size-dependent failure, governed by the limited spatial reach of Mn-assisted redox stabilization, provides a direct atomistic explanation for the irreversible capacity loss observed in deeply cycled Li-rich cathodes.

\subsection{Kagome-like Mn network enables high capacity reversibility}

Our analysis reveals that the topology of the Mn network plays an important role in determining the structural and capacity reversibility Li$_2$MnO$_3$. The conventional honeycomb Mn lattice is topologically vulnerable; its sparse connectivity allows uncontrolled (interlayer) Mn migration during charging, which promotes the growth and coalescence of nanovoids. Once these voids exceed the repair threshold ($>1$ nm), they can no longer be healed during discharge, leading to irreversible structural damage and permanent capacity loss.

\begin{figure}[htb]
\centering
\includegraphics[width=17 cm]{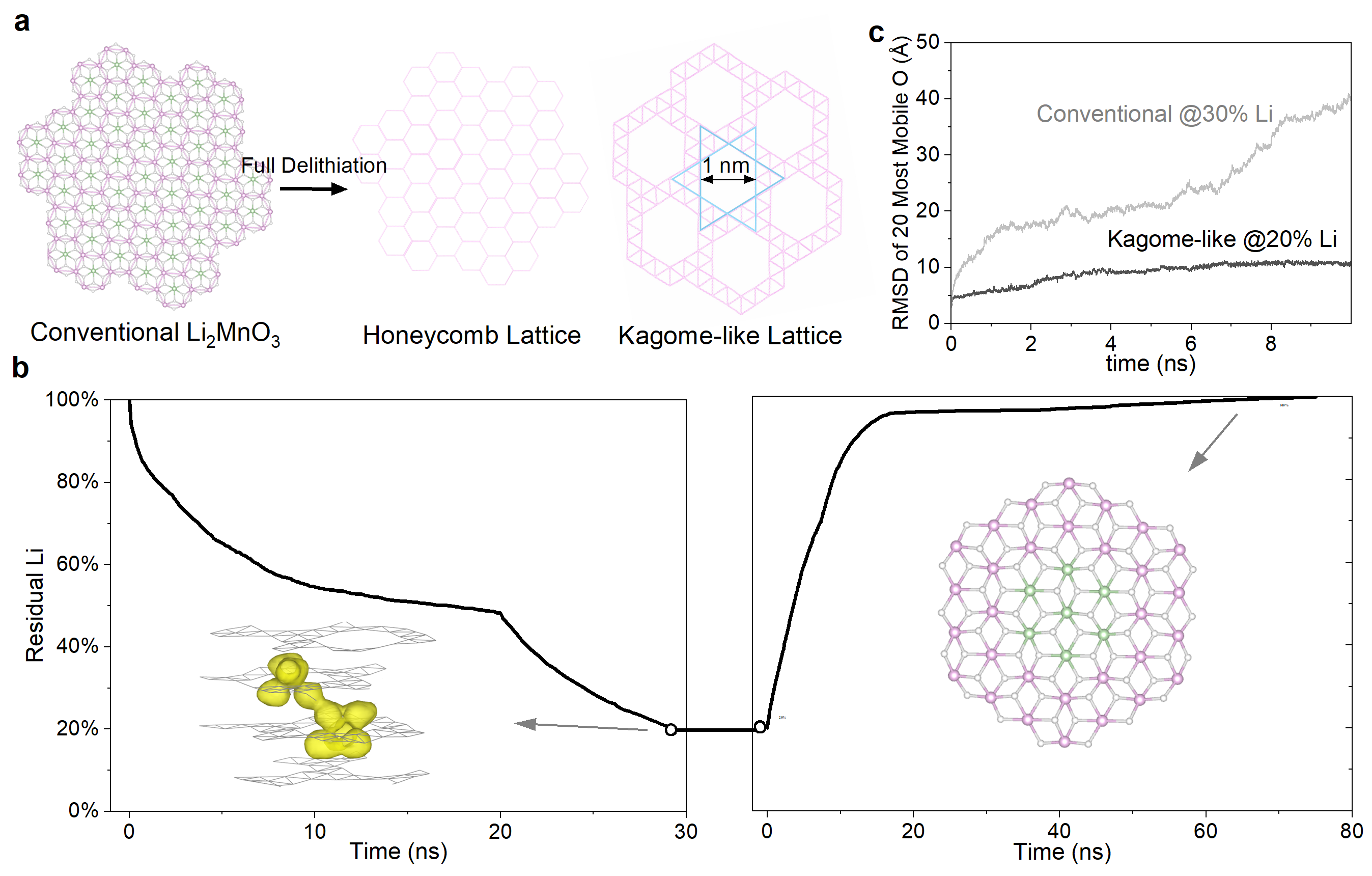}
 \caption{High reversible capacity and structural stability of Kagome-like Li$_2$MnO$_3$.
 (a) Mn network transformation via topological engineering. 
 The conventional honeycomb Mn lattice is reconfigured into a Kagome-like framework composed of triangular Mn clusters. This topology introduces a regular array of $\approx$1 nm voids in the fully delithiated state.
 (b) Near-perfect reversibility of Kagome-like Li$_2$MnO$_3$ under 80\% delithiation. The inset shows O$_2$ molecules remain trapped within pre-designed voids.
(c) RMSD analysis of the 20 most mobile oxygen atoms in conventional and Kagome-like electrodes at 20\% residual Li. 
}
\label{fig_super}
\end{figure}

Here, rather than mitigating structural damage post hoc, we propose a new design paradigm: preventing degradation through rational topological engineering of the Mn network. We computationally designed a novel structural motif for Li$_2$MnO$_3$ by reconfiguring the Mn framework, as illustrated in Fig~\ref{fig_super}a. The conventional honeycomb lattice was transformed into a Kagome-like arrangement composed of densely connected triangular Mn clusters. This new topology inherently features a regular array of $\approx$1 nm nanovoids, which are fully occupied by Li and O atoms in the stoichiometric Li$_2$MnO$_3$ composition. 

To validate this topological design, we performed a full charge-discharge simulation, pushing the engineered structure to an aggressive 80\% delithiation (20\% residual Li).  As shown in Fig.~\ref{fig_super}b, the system exhibits nearly 100\% capacity recovery. The inset confirms complete structural restoration upon full relithiation. We compared the RMSD of the oxygen atoms under deep delithiation ( Fig.~\ref{fig_super}c). Specifically, the conventional topology with 70\% delithiation fails catastrophically, with a runaway increase in oxygen RMSD indicating void percolation and the formation of an O$_2$ escape channel. In contrast, the engineered topology remains stable even at 80\% delithiation, effectively suppressing O$_2$ mobility. The oxygen RMSD stays low and stable, demonstrating that O$_2$ molecules remain confined within their pre-designed 1 nm cages..

\section{Conclusion}

In this work, we elucidate the fundamental atomic-scale mechanisms that govern oxygen redox reversibility in Li-rich cathodes represented by Li$_2$MnO$_3$. Through submicrosecond-scale molecular dynamics simulations with first-principles accuracy, we move beyond static pictures of degradation to reveal a dynamic interplay between void evolution and the underlying transition metal framework. We show that the conventional honeycomb structure is topologically vulnerable, leading to two primary failure mechanisms. The first involves the uncontrolled growth and percolation of O$_2$-filled voids, driven by a self-catalytic mechanism and facilitated by stochastic interlayer Mn migration. The second is a size-dependent irreversibility: voids exceeding $\approx$1 nm in diameter become unrecoverable upon discharge, as their cores extend beyond the catalytic influence of the surrounding Mn network.

These findings point to a design paradigm that shifts the focus from mitigating damage to engineering inherent stability. We propose that the topology of the Mn network is the key variable controlling structural integrity and electrochemical reversibility. Based on this topology-centric design principle, we computationally engineered a novel structure of Li$_2$MnO$_3$ with Kagome-like Mn sublattice. This structure is composed of a  network of dense, triangular Mn clusters that act as a rigid scaffold, pre-defining a regular array of subnanometer voids. This engineered topology successfully cages O$_2$ molecules, preventing void coalescence and percolation even at an aggressive 80\% delithiation, thereby ensuring the voids remain within the critical size limit for complete repair. As a result, our design demonstrates full structural and electrochemical reversibility. The realization of such an ordered Kagome-like cation arrangement, while challenging, is a tangible goal for advanced synthesis protocols such as topochemical reactions or controlled ion exchange~\cite{House20p502}. By demonstrating that atomic-level topological control can overcome the long-standing stability issues of anion-redox materials, we open a pathway toward high-energy-density batteries with exceptional cycling performance.

\newpage
\begin{acknowledgments}
C.K., Y.Y., and S.L. acknowledge the support from Zhejiang Provincial Natural Science Foundation of China (LR25A040004). J.W. acknowledges the support from Natural Science Foundation of China (21975207). The computational resource is provided by the Open Source Supercomputing Center of S-A-I and Westlake HPC Center. 
\end{acknowledgments}

\bibliography{L73}
\newpage
\clearpage
\newpage

\clearpage
\newpage

\clearpage
\newpage

\clearpage
\newpage
\begin{figure}[htb]
\centering
\includegraphics[width=17 cm]{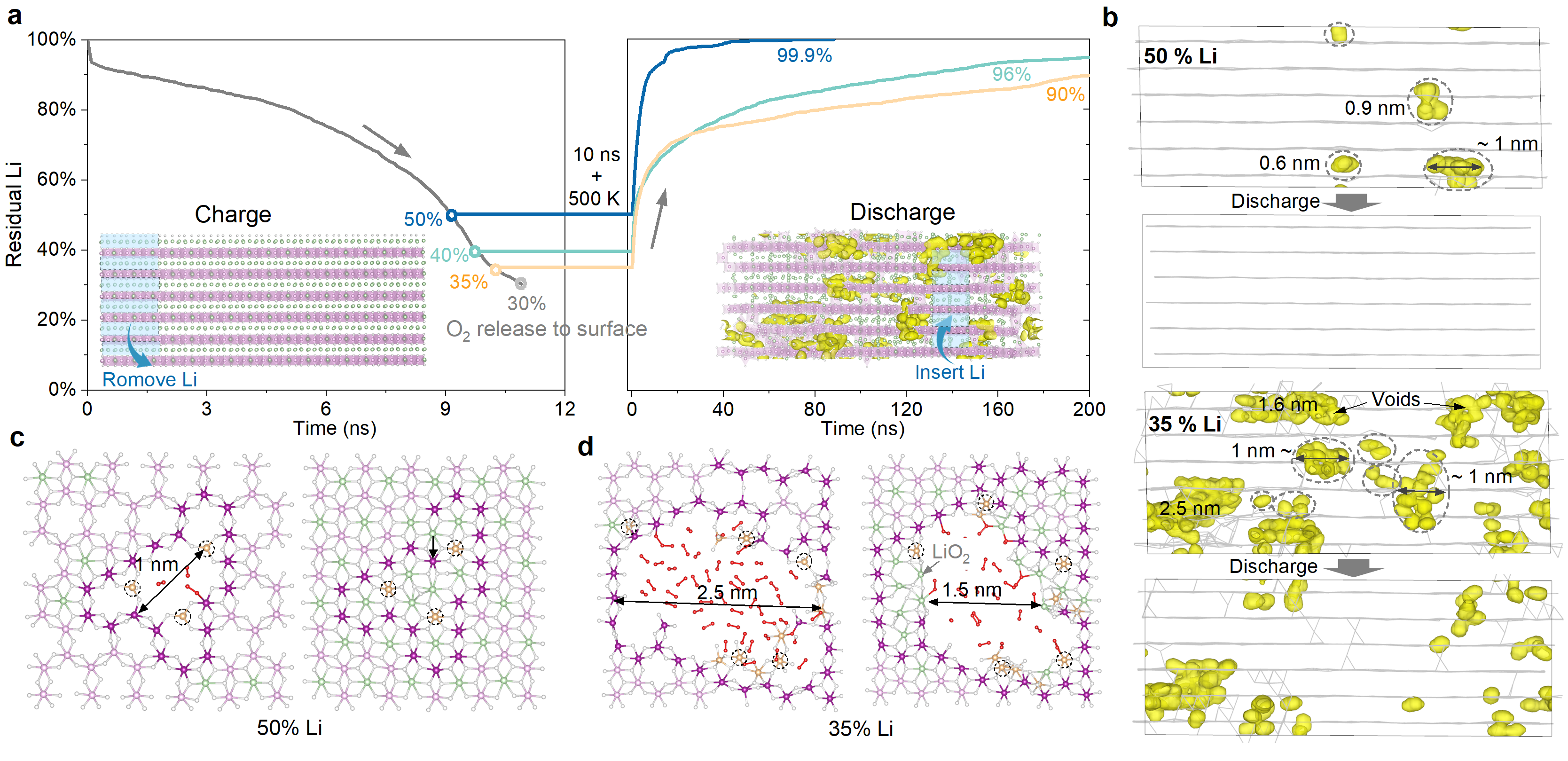}
 \caption{MD simulations of full charge-discharge cycles and void repair behavior at different charge states.
(a) Charge--discharge simulation protocol.
Schematic of the full-cycle simulation using DPMD. Structures with 50\%, 40\%, and 35\% residual Li were equilibrated and used as starting points for discharge simulations, during which Li atoms were progressively reinserted into a designated region of the supercell.
(b) Post-discharge void distributions. At 50\% residual Li, small voids ($<1$ nm) were fully repaired. At 35\% residual Li, larger voids ($>1$ nm) remained partially unrepaired, correlating with irreversible capacity loss in (a).
(c) Complete void repair at 50\% residual Li.
A small $\approx$1 nm void fully heals during discharge. Mn ions near the void wall assist in stabilizing LiO$_2$ intermediates. Those exhibiting significant out-of-plane displacements are highlighted in gold.
(d) Incomplete void repair at 35\% residual Li.
A large $\approx$2.5 nm void contracts to $\approx$1.5 nm but remains unrepaired. The void core falls outside the catalytic range of Mn, preventing further redox activity and trapping O$_2$.
 }
\label{fig_discharge}
\end{figure}
\clearpage
\newpage

\end{document}